\documentstyle[epsfig]{mn}

\newcommand{\etal}{{et al.\ }}
\newcommand{\mnras}{MNRAS}
\newcommand{\apj}{ApJ}

\title{The Milky Way's satellite population in a $\Lambda$CDM universe}
\author[F. Stoehr, S.~D.~M. White, G. Tormen, V. Springel]
{Felix Stoehr$^1$, Simon D.~M. White$^1$, Giuseppe Tormen$^2$, 
Volker Springel$^1$\\
$^1$Max-Planck-Institut f\"{u}r Astrophysik, Karl-Schwarzschild-Str. 1, 
85748 Garching bei M\"unchen, Germany\\
$^2$Dipartimento di Astronomia, Universita di Padova, vicolo 
dell'Osservatorio 5, 35122 Padova, Italy\\
\smallskip
Email: felix@mpa-garching.mpg.de, swhite@mpa-garching.mpg.de, 
tormen@pd.astro.it, volker@mpa-garching.mpg.de}
\date{MNRAS, accepted, July 23, 2002}

\begin{document}
\maketitle

\begin{abstract}
We compare the structure and kinematics of
the 11 known satellites of the Milky Way with high resolution 
simulations of the formation of its dark halo in a $\Lambda$CDM
universe. In contrast to earlier work, we find excellent agreement. 
The observed kinematics are exactly those predicted for stellar 
populations with the observed spatial structure orbiting 
within the most massive ``satellite'' substructures in our simulations. 
Less massive substructures have weaker
potential wells than those hosting the observed satellites.
If there is a halo substructure ``problem'', it consists
in understanding why halo substructures have been so inefficient in
making stars. Suggested modifications of dark matter properties
(for example, self-interacting or warm dark matter) may well
spoil the good agreement found for standard Cold Dark Matter. 
\end{abstract}


\section{Introduction}

In hierarchical cosmologies such as the now-standard $\Lambda$CDM
model, objects like the dark halo of our Milky Way grow
through the merging of previously collapsed systems with
a wide range of masses. Even the earliest detailed models for the 
growth of a ``Milky Way'' showed that simple assumptions for
star formation efficiency imply many more visible 
satellites than are actually observed (Kauffmann, White \& Guiderdoni 
1993). These authors argued that gas condensation and thus star formation  
must be strongly suppressed in systems with low escape velocity,
perhaps by photoionisation heating. This echoed Efstathiou's
(1992) arguments on the related question of why the faint-end 
slope of the observed galaxy luminosity function is much shallower 
than predicted by the simple hierarchical galaxy formation theory 
of White \& Rees (1978) and its later reworkings in the CDM context
(e.g. White \& Frenk 1991).

This issue drew relatively little attention until N-body techniques 
became capable of simulating halos with hundreds of thousands of 
particles. Moore \etal (1999) and Klypin \etal (1999) showed that 
galaxy halos in a CDM cosmogony are not smooth
systems. Roughly 10\% of the mass within their virialised regions
is contributed by a host of dense self-bound substructures.  These are
the surviving cores of objects which merged together to make the final 
system, and so correspond directly to the overabundant satellites of
the earlier models. The apparent overabundance may again be
reconciled with the small number of visible satellites by
invoking the inhibiting effects of photo-heating
(Bullock, Kravtsov \& Weinberg 2000, Benson \etal 2002). Both Moore \etal
(1999) and Klypin \etal (1999) emphasised a different
problem, however. The maximum circular velocities of the 10 or 20
most massive substructures within a Milky Way halo are predicted to be
in the range 20 to 60 km/s, whereas the observed velocity dispersions
of 7 of the Milky Way's 11 satellites are below 10 km/s. Halo models
typically have well over a hundred substructures with maximum circular 
velocity above 10 km/s.

This discrepancy has been considered a ``crisis'' for the conventional
$\Lambda$CDM cosmogony, prompting proposals to modify the microscopic 
properties of the dark matter (e.g. Spergel \& Steinhardt 2000; Moore 
\etal 2000; Yoshida \etal 2000; Bode, Ostriker \& Turok 2001). We here 
examine the discrepancy more critically, using $\Lambda$CDM
simulations with similar resolution to the best simulations of Moore 
\etal (1999) or Power \etal (2002). We analyse the potential well 
structure of the most massive ``satellites'' in our final system,
calculating the velocity dispersion profiles predicted if stellar
systems identical in structure to the observed satellites are placed
in each. Remarkably, we find excellent agreement with the observed
velocity dispersions, even for systems like Fornax and Draco where
resolved profiles are available (Mateo 1997; Kleyna \etal 2002).

As we were completing this work, Hayashi \etal (2002) posted a preprint 
addressing similar issues (and with similar conclusions). Their
approach is complementary to the one we adopt here. They do not analyse 
a fully consistent simulation of a Milky Way halo and its
substructure, nor do they compare in detail with the 
kinematic structure of the observed dwarfs. On the other hand, they
carry out a much more complete and reliable analysis of the structural
evolution of individual satellite substructures than is possible
with our own simulation data. We will use their results below to 
provide an independent check on the most uncertain aspect of our 
own simulations -- whether they give reliable estimates for the 
inner ``core'' structure of satellite subhalos.

In the next section we will give a brief description of the simulations
we have carried out, the methods we have used to find substructures
and to characterise their potential wells, and the checks we have made
on the reliability of these measures. Section 3 then shows how to
predict the line-of-sight velocity dispersion profile for a dwarf
galaxy of given spatial structure within a given potential well. We
apply the technique to the 11 known satellites of the Milky Way
and the 20 deepest satellite potential wells of our highest resolution 
simulation. Section 4 discusses our results, their principal remaining
uncertainties and their implications.

\section{The $N$-body simulations}

We work with a flat $\Lambda$-dominated Cold Dark Matter universe,
with matter density $\Omega_{\rm m}=0.3$, cosmological constant 
$\Omega_{\Lambda}=0.7$, expansion rate $H_{0}=70~$km s$^{-1}$Mpc$^{-1}$, 
index of the initial fluctuation power spectrum $n=1$, and present-day
fluctuation amplitude $\sigma_8=0.9$. We begin with a
dark matter simulation of a ``typical'' region 
of the Universe ($N\sim 6\times 10^7$, particle mass 
$\sim 2\times 10^8M_\odot$) for which the techniques of Springel \etal (2001b,
hereafter SWTK) have been used to follow the formation of the galaxy 
population (Stoehr \etal 2002, in preparation). We identify a
relatively isolated ``Milky Way like'' galaxy and resimulate its 
halo at a series of higher resolutions, again using techniques
from SWTK and the $N$-body code {\sc gadget} (Springel, Yoshida \& White
2001a). In the simulations GA0, GA1 and GA2 analysed in this paper 
the resimulated halo has 13603, 123775 and 1055083 particles respectively 
within $r_{200}$, the radius enclosing a mean
density 200 times the critical value. The density profiles of these
three resimulations agree well and are accurately fit by the NFW
formula with concentration $c=10$ (Navarro, Frenk \& White 1997). We
show the corresponding circular velocity curves in Figure 1a. These
have been scaled down by a factor of 0.91 in velocity and radius 
(corresponding to a factor of 0.74 in mass, but unchanged density
and time scales) so that they peak at 220~km/s. With this scaling,
dark matter particle masses are $1.8\times10^8$, $1.9\times10^7$ 
and $2.0\times10^6 M_\odot$, and Plummer equivalent softening lengths 
are 1.8, 1.0 and 0.49 kpc in GA0, GA1 and GA2, respectively. In all 
three $r_{200}\approx 270~$kpc. Note that since the Milky Way's stars 
contribute significantly to its measured rotation velocity, our chosen 
scaling probably produces too large a mass for the Milky Way's halo and 
thus also for substructures within it. This is conservative for the 
purposes of this paper.

We identify self-bound substructures within our final halos using
the procedure {\sc subfind} described in detail by SWTK. In Figure 1b we
compare the number of substructures found within $r_{200}$ in each of 
our simulations. For each we plot cumulative abundances down to 
a mass corresponding to 20 particles. Clearly the agreement is good. 
For example, down to $4\times 10^9M_\odot$, the 20 particle limit for
GA0, we find 5, 7 and 2 subhalos in GA0, GA1 and GA2, respectively,
well within the fluctuations expected for Poisson statistics. Down to
the 20 particle limit for GA1 ($4\times 10^8M_\odot$) there are 28 and
27 subalos in GA1 and GA2. We have also 
verified that our cumulative function of peak velocities is consistent 
with that of Font \etal (2001). 

For each subhalo we define the centre 
as the position of the densest particle, where densities are evaluated 
using an SPH smoothing kernel. We then calculate a circular velocity curve 
as $V_c(r) = (GM(r)/r)^{1/2}$ where $M(r)$ is the total mass within $r$ of 
this centre. In almost all of the more massive subhalos, these curves rise 
from the centre, peak at a distance of a few kpc, and then fall for a 
while before eventually rising again because of the contribution from 
the smooth halo. Fitting such curves we can define a maximum circular 
velocity and a radius at which it is achieved for each subhalo. In 
Figure 1c we plot these two quantities against each other for all 
subhalos in GA1 and GA2 with more than 100 and 300 particles, respectively. 
There is a weak correlation but no clear tendency for the GA1 subhalos 
to be less concentrated (i.e. to have larger $r_{\rm max}$ at given 
$V_{\rm max}$) than those in GA2. By comparing the evolution of the
two simulations in detail it is possible to identify counterparts in 
GA2 for all the eight GA1 subhalos plotted in Figure 1c. Four of
these lie within $r_{200}$ in GA2, and the corresponding pairs of
points are indicated with ellipses in the plot. There is good
agreement between the $(V_{max},r_{max})$ values measured in the two
simulations despite the factor of 10 difference in mass resolution.

In Figure 1d we plot circular velocity curves for some of the 20 GA2 
subhalos with the largest values of $V_{\rm max}$. Even the least massive 
of these has more than 300 particles, and we have checked that they all
have maintained their structure with relatively little change since
at least $z=0.4$. It is noticeable that these circular velocity
curves have narrower peaks than the NFW form, so we have
fitted them with parabolae
\begin{equation}
\log (V_c/V_{\rm max}) = -a \, [\log(r/r_{\rm max})]^2,
\end{equation}
adjusting the constant $a$ to get a good fit at $r<r_{\rm max}$. We find 
values for $a$ ranging from 0.25 to 0.7 with a median of 0.45. These
fits are shown as dotted curves in Figure 1d, and we use them
to represent the circular velocity data in the analysis of the following 
section. This analysis depends quite sensitively on the shape of
the curves at small radius. This is a matter for concern since in this
regime, our results are likely to be affected by discreteness effects and by 
gravitational softening. Although it is reassuring that we see little 
temporal evolution in the inner structure of individual subhalos, and that
the more massive GA1 halos agree in structure with their GA2
counterparts, a more convincing demonstration that equation (1) is 
realistic comes from the fact that it fits the circular
velocity curves of the much higher resolution simulations of stripped 
satellites by Hayashi \etal (2002). The curves in their Figure 10 are 
well fit by values of $a$ in the range 0.37 to 0.65. Below we will repeat 
our analysis using $a=0.37$ for all our substructures, rather than the 
individual values estimated from their detailed internal structure. This 
choice maximises the circular velocity at the relevant radii ($r\sim
0.1 r_{\rm max}$) for curves compatible in shape with those of
Hayashi \etal and for the values of $V_{\rm max}$ and $r_{\rm max}$ measured
in our simulation.

\begin{figure*}
\centering
\epsfig{file=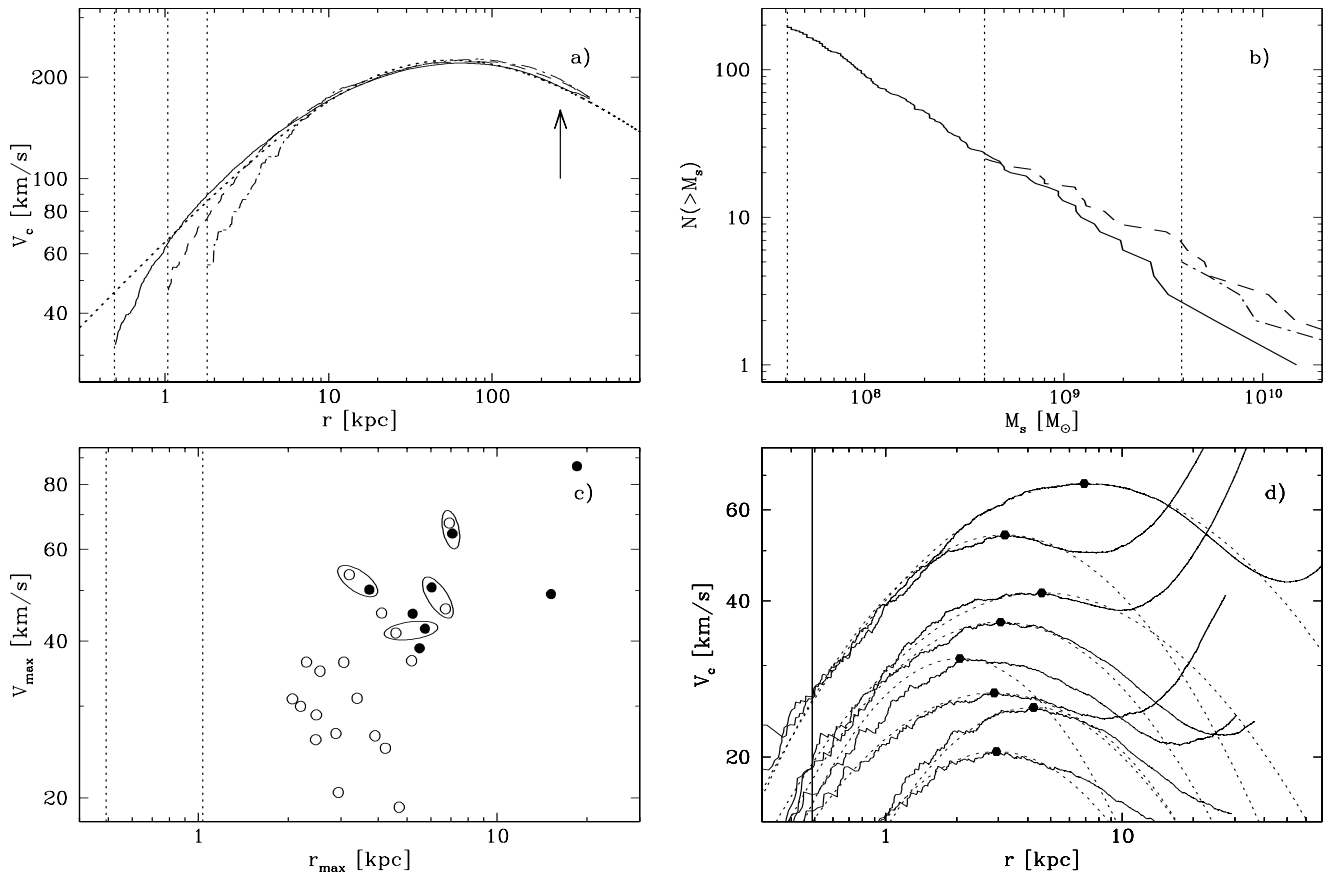,width=\textwidth}
\caption[Properties of simulated halos and their substructure]
{(a) Circular velocity curves for the ``Milky Way'' halos
in our three resimulations, scaled to peak at 220 km/s. 
An arrow indicates $r_{200}$. We overplot a NFW fit with $c=10$
(dotted line). Vertical lines show the Plummer softening
lengths. (b) The cumulative number of
subhalos within $r_{200}$ as a function of their mass for these same
three resimulations. Vertical lines show the 20 
particle limits. (c) Maximum of the circular velocity 
curve against the radius at which it is reached for all subhalos with more 
than 100 particles in GA1 (filled circles) and with more than 300 
particles in GA2 (open circles). Vertical lines show the softening 
lengths and ellipses enclose counterpart subhalos. (d) Circular
velocity curves and fits to equation (1) for subhalos 1, 2, 5, 8, 11,
14, 17 and 20 in GA2 in order of decreasing $V_{\rm max}$. 
The vertical line shows the softening length.}
\end{figure*}

\section{Subhalo potentials and the observed Milky Way satellites}

\begin{table}
\caption{Core radii $r_c$, tidal to core radius ratios $r_t/r_c$ and central 
velocity dispersions $\sigma_0$ for the Milky Way dwarf spheroidals. The last 
two columns show the number of simulated subhalos for which the predicted central 
velocity dispersion is larger than that observed. The first takes $a$ in equation 
(1) from a fit to our data, the second fixes $a=0.37$.}
\begin{center}
\begin{tabular}{l l c c c c}
   \hline 
    & $r_c$ [kpc] & $r_t/r_c$  & $\sigma_0$ [$\frac{km}{s}$] 
       & N$_{GA2}$ & N$_{Hay}$ \\    
    \hline
    Sagittarius & 0.44  & 6.8 & 11.4(19) & 11(2) & 15(1)\\
    Fornax      & 0.46  & 5.1 & 10.5     & 13    & 15\\
    Leo I       & 0.215 & 3.8 & 8.8      & 4     &  5\\
    Sculptor    & 0.11  & 13  & 6.6      & 4     &  5\\    
    Leo II      & 0.16  & 3   & 6.7      & 1     &  5\\    
    Sextans     & 0.335 & 9.6 & 6.6      & 18    &  19\\    
    Carina      & 0.21  & 3.3 & 6.8      & 6     &  10\\    
    Ursa Minor  & 0.20  & 3.2 & 9.3      & 0     &  2\\        
    Draco       & 0.18  & 5.2 & 9.5      & 0     &  2\\            
    \hline    
\end{tabular}
\end{center}
\end{table}

Except for the Magellanic Clouds, all the known satellites of the
Milky Way are dwarf spheroidal galaxies. All have measured
core and tidal radii, as well as measured central line-of-sight 
velocity dispersions. We summarise the data in Table 1. They are
taken primarily from the review of Mateo (1998), but we have preferred
the more recent density profile parameters measured for Draco by 
Odenkirchen \etal (2001). Sagittarius is also a special case. 
Here we take the core and tidal radii from the {\it pre-disruption} 
model of Helmi \& White (2001) and carry out our analysis
using both the current central velocity dispersion given by Mateo 
and the higher pre-disruption value from the model of Helmi \&
White. The latter value and quantities based on it are
indicated in parentheses in Table 1. For Fornax and Draco there are 
also published velocity dispersion profiles with good signal-to-noise;
we compare these to our $\Lambda$CDM predictions below. To 
characterise the potentials of the Magellanic Clouds we take 
a circular velocity of 50 km/s at a distance of 5 kpc for the LMC 
(van der Marel \etal 2002) and a circular velocity of 60 km/s at a 
distance of 2.5 kpc for the SMC (Stanimirovic 2000). 
Comparing with Figure 1d, we see that the Clouds fit well in 
the two most massive subhalos of GA2.
We now ask whether the next ten or so most massive simulated
subhalos could host the observed dwarf spheroidals. 

A spherical stellar system with number density profile $\rho(r)$
in equilibrium within a spherical potential well with circular
velocity curve $V_c(r)$ will, if its velocity dispersion tensor
is everywhere isotropic, satisfy
\begin{equation}
{{\rm d}(\rho\sigma^2)\over {\rm d}r}= -{\rho V_c^2\over r},
\end{equation}
where $\sigma(r)$ is the one-dimensional velocity dispersion profile.
Assuming that $\rho(r)$ goes to zero at the finite ``tidal'' radius 
$r_t$, this equation can be integrated to obtain 
\begin{equation}
\rho(r)\sigma^2(r) = \int_r^{r_t} {\rm d}r^\prime~\rho V_c^2/r^\prime .
\end{equation}
Projecting $\rho\sigma^2$ and $\rho$ and taking their ratio, we can
obtain $\sigma_p(r_p)$, the line-of sight velocity dispersion
at projected distance $r_p$ from the satellite centre,
\begin{equation}
\sigma_p^2(r_p) = {\int_{r_p}^{r_t} {\rm d}r~\rho V_c^2 (r^2 -
r_p^2)^{1/2}/r \over \int_{r_p}^{r_t} {\rm d}r~\rho r/(r^2 - r_p^2)^{1/2}}. 
\end{equation}
We apply this formula to stellar systems with the observed
structure of the dwarf spheroidals embedded in the potential
wells of our simulated subhalos. We will represent the latter
by equation (1) and the former by King's (1962) fitting formula
\begin{equation}
\rho(r)\propto \left[\arccos(z)/z - (1-z^2)^{1/2}\right]/z^2,
\end{equation}
where
\begin{equation}
z^2 = {1 + r^2/r_c^2\over 1 + r_t^2/r_c^2}.
\end{equation}

We have used this method to calculate the expected central
line-of-sight velocity dispersion of a stellar system with the
core and tidal radii of each of the observed dwarf spheroidals
embedded in each of the 20 most massive substructures in our
GA2 halo. For each dwarf spheroidal, we count the number of 
subhalos for which the predicted central dispersion
is larger than that observed, and we record this number in Table 1.
We have carried out this exercise twice; once using the fits to
the individual simulated circular velocity curves (see Figure 1d) and
once taking $r_{\rm max}$ and $V_{\rm max}$ from the simulation and then using
$a=0.37$ in equation (1) for all subhalos. The latter procedure forces 
the inner structure of each subhalo to resemble that of the most
concentrated of the stripped 
satellites simulated by Hayashi \etal (2002). These numbers show a 
remarkable result. For both assumptions all 11 of the Milky Way's 
satellites can be accommodated within the 20 most massive subhalos, 
and all but 3 can be accommodated within the 10 most massive (all but
2 if the higher dispersion value is adopted for Sagittarius). Given
that substantial realisation to realisation fluctuations are expected 
in the properties of the more massive subhalos, there is 
surprisingly good agreement between the kinematics of the observed
satellites and those predicted by our $\Lambda$CDM simulation.

\begin{figure}
\centering
\epsfxsize=\hsize\epsffile{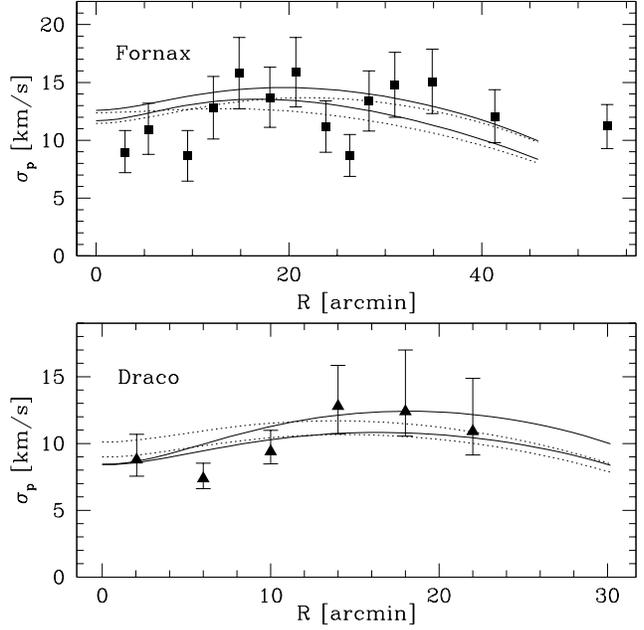}
\caption[Velocity dispersion profiles for Fornax and Draco]
{Observed velocity dispersion profiles for Fornax and for Draco
are compared with those predicted for stellar systems with the observed
density structure embedded in some of the most massive
dark matter satellites in our $\Lambda$CDM simulation. Solid
lines show results for circular velocity curves estimated
directly from our simulation (GA2), while dotted lines assume a 
circular velocity curve shape similar to that of the most concentrated 
satellites in the simulations of Hayashi \etal (2002).
}
\end{figure}

For Fornax and Draco, more detailed kinematic data have been published
and allow a closer comparison. In Figure 2 we reproduce
velocity dispersion profiles from Mateo (1997) for Fornax, and from
Kleyna \etal (2002) for Draco, and we overplot the predictions of the
above equations for a number of our more massive subhalos. From this
figure it is clear that GA2 not only predicts the correct central
velocity dispersion values, but also the correct shapes for the
dispersion profiles. Within the observational uncertainties this is 
equally true for circular velocity curves taken directly from GA2
and for curves with a the most concentrated shape consistent with 
the simulations of Hayashi \etal (2002). The maximum circular
velocities of the subhalos compared with the Fornax data range 
between 31 and 46 km/s. The corresponding range for the subhalos 
compared with Draco is 35 to 54 km/s. In a $\Lambda$CDM model the 
dark halos of dwarf spheroidals are predicted to have maximum circular 
velocities much larger than the observed velocity dispersions. 
It is a curious that the observational data require Draco and Fornax 
to have similar halos despite the fact that Fornax is more luminous 
by a factor of almost 60.
 
\section{Discussion}

The last section shows that the potentials of
the most massive subhalos in our $\Lambda$CDM simulation are in 
excellent agreement with the observed kinematics of the Milky 
Way's satellites. There is some reason to be cautious, because subhalo
concentrations could be significantly underestimated as a result of 
numerical limitations, in particular the relatively large
gravitational softening and particle mass in even our highest 
resolution simulation. We have argued from a comparison of simulations 
with differing resolution that the induced bias in the values of
$V_{\rm max}$ and $r_{\rm max}$ is small for objects as massive as
those which concern us here. Furthermore, our subhalo 
circular velocity curves agree well at small radii with those 
which Hayashi \etal (2002) obtain from {\it much} higher resolution 
simulations of the tidal stripping of individual satellites. These 
authors explain why tidal effects produce objects with {\it lower} 
central concentration than isolated halos of similar maximum circular 
velocity. Even adopting the most concentrated profile consistent
with their simulations has only a minor effect on our analysis. 
(Compare the rankings for the two cases in Table 1.) On the basis 
of the currently available observational and simulation data, it 
seems more appropriate to consider
the observed kinematics of the Milky Way's satellites as a 
triumph for the $\Lambda$CDM model than as a crisis.

If the observed dwarf spheroidals do indeed have dark halos of the
kind that both we and Hayashi \etal (2002) suggest, there are a 
number of interesting consequences. In the first place, the measured 
``tidal radii'' can have nothing to do with tidal effects, but must 
reflect an edge to the visible stellar population within a much more 
extended dark halo. This kind of structure appears to be a direct and 
inevitable consequence of the flat or rising velocity dispersion profiles 
measured in Fornax and Draco. Tidal tails and extra-tidal stars should not, 
then, be present in most systems. Although there is no problem accommodating 
a single disrupting object like Sagittarius, it would become uncomfortable
if tidal stripping were detected unambiguously in other systems. 
There are some indications that Carina might be such a case 
(Majewski \etal 2000).

A second consequence is that the total mass associated with the
dwarf spheroidals is much larger than usually assumed.
The tenth most massive subhalo in our GA2 simulation has a bound
mass of $1.2\times 10^9M_\odot$ and its progenitor system was several
times more massive just before it fell into the Milky Way's
halo. This sharpens the problem in understanding why the dwarf
spheroidals have formed stars with such low efficiency; their
stellar masses lie in the range $10^6$ to $10^8M_\odot$. 
Clues presumably lie in their comparatively narrow
ranges of size and velocity dispersion, and in
their surprisingly varied star formation histories (Mateo 1998).

Finally, if the observed satellites do occupy the most
massive subhalos in a $\Lambda$CDM model, then 
many smaller subhalos are presumably also present but are devoid of 
stars. It is interesting to look for observable 
consequences of their existence. Dynamical effects, for example
heating or distortion of the Galactic disk or scattering of the orbits
of visible halo objects, are weak, and are dominated by the most massive
subhalos rather than by their more abundant low mass brethren. 
Such effects may, nevertheless, be detectable in favorable cases (Font 
\etal 2001; Johnston, Spergel \& Haydn 2001). It may also be possible to
detect substructure in gravitational lens galaxies through the 
statistics of flux ratios in samples of multiply imaged quasars 
(Mao \& Schneider 1998; Chiba 2002;  Dalal \& Kochanek 2002);
again the effects are 
dominated by the few most massive subhalos. If dark matter 
detection experiments are successful, it may become feasible to 
search for structure in the dark matter distribution at the Earth's 
position. Detailed analysis suggests that in a $\Lambda$CDM model 
the signal will be difficult, but perhaps not impossible,
to see (Helmi, White \& Springel 2002).

The fact that the core structure of satellite subhalos in a
$\Lambda$CDM model agrees with the kinematics of observed 
satellites is {\it prima facie} evidence against dark matter
with modified properties, for example, Warm Dark Matter or
Self-Interacting Dark Matter. In recent work, the primary motivation
for considering such modifications to the microscopic physics of
the dark matter particles has been to reduce the concentration of halos
and the abundance of substructure (e.g. Spergel \& Steinhardt 2000;
Yoshida \etal 2000; Bode \etal 2001). A significant reduction of the
central concentration of satellite subhalos in our simulations would,
however, make it difficult to produce satellites with velocity 
dispersions {\it as large} as those observed. 

%\acknowledgments 

\end{document}